\documentclass[aps,pra,twocolumn,preprintnumbers,amsmath,amssymb]{revtex4}

\usepackage{graphicx}
\usepackage{dcolumn}
\usepackage{bm}

\newcommand{\bra}[1]{\langle#1| }
\newcommand{\ket}[1]{|#1\rangle }

\begin{document}


\title{Inhomogeneous Quantum Walks}

\author{Noah Linden}
 \email{n.linden@bristol.ac.uk}
 \affiliation{Mathematics Department, University of Bristol}

\author{James Sharam}
 \email{james.sharam@bristol.ac.uk}
 \affiliation{Department of Mathematics, University of Bristol, Bristol BS8 1TW, United Kingdom}

\date{19th June 2009}

\begin{abstract}
We study a natural construction of a general class of inhomogeneous quantum walks (namely walks whose transition
probabilities depend on position).  Within the class we analyze walks that are periodic in position and show
that, depending on the period, such walks can be bounded or unbounded in time; in the latter case we analyze the
asymptotic speed.  We compare the construction to others in the existing literature.  As
an example we give a quantum version of a non-irreducible classical walk: the P\'{o}lya Urn.
\end{abstract}

\maketitle

\section{\label{sec:section1} Introduction}
The study of classical random walks on a lattice has a long history and numerous applications in fields such as simulation of physical processes and probability theory. A random walk starts with a particle at a node on a lattice, then at each time step the particle jumps to another node with a given probability. Inhomogeneous random walks are those that that have position and direction dependent probabilities of jumping to neighbouring nodes.

Quantum walks \cite{aharonov1993,meyer1996a,meyer1996b,watrous2001} have been shown to have many applications in quantum algorithms, such as algorithms for searching \cite{shenvi2003},  for the element distinctness problem \cite{ambainis-2007-37}, for matrix product verification \cite{buhrman2004}, for testing group commutativity \cite{magniez-nayak2007} and for triangle finding \cite{magniez2007b}. Quantum interference causes quantum walks to behave in a qualitatively different way from their classical counterparts. Most of the original work considered homogeneous walks (where the amplitude for moving does not depend on position).  In this paper we analyze a natural construction
of inhomogeneous walks and, for a simple family of such walks, show that they can be bounded or unbounded in time.  We have  a  number of motivations for this work:
 to gain an understanding of the most appropriate general setting for quantum walks, to probe the possible long time behaviours of such walks, and also the longer term goal of producing techniques that may be useful for producing quantum algorithms.

The idea of looking at inhomogeneous walks is not new.  For example the recognition that it
 is natural to allow coins to be position dependent may be found in \cite{ambainis2003,ambainis2005,santha2008}.  In \cite{szegedy} an alternative general method is given for quantizing a classical Markov chain.  The method (particularly as generalised in \cite{magniez2007a}) gives a large class of inhomogeneous walks.  However as we shall see later, the construction in this paper is not equivalent to that in \cite{szegedy,magniez2007a}.  As an example we give a quantum analogue of a reinforced process: the P\'{o}lya Urn.  In much of the earlier part of the paper we focus on walks on the line, but it should be clear (and the quantum P\'{o}lya Urn is an example) that our construction is applicable to general graphs.

For comparison with what comes later, we briefly review a simple homogeneous walk on the line.
Let $ \{ \ket{n} : n \in \mathbb{Z} \} $ be an infinite set of states where each represents the position $n$ along the infinite line of integers. In addition there is a coin register that is spanned by two states  ($\ket{L}$ and $\ket{R}$) which represents the direction of motion at a particular time step. The evolution of the quantum walk proceeds by applying a coin operator to the coin states, in order to select which direction to move in with a certain probability amplitude, and then a shift operator to move the resulting amplitudes along the line to their new position. The most commonly used coin operator $C$ is the Hadamard operator, corresponding to a 50\% chance of moving left or right, and in this basis is given by:
\begin{equation}
C = \frac{1}{\sqrt{2}} \Big( \ket{L} + \ket{R} \Big) \bra{L} + \frac{1}{\sqrt{2}} \Big( -\ket{L} + \ket{R} \Big) \bra{R}
\end{equation}
The most commonly used shift operator $S$, which we will be used throughout this paper is:
\begin{equation}
\label{standard-shift}
S = \displaystyle \sum_n \ket{n-1,L}\bra{n,L} + \ket{n+1,R}\bra{n,R}
\end{equation}
A full time step $W$ of this quantum random walk is therefore:
\begin{equation}
W = S(I \otimes C)
\end{equation}
So the state $\ket{\Psi}$ of the walk after $T$ time steps is therefore:
\begin{equation}
\ket{\Psi(T)} = W^T \ket{\Psi(0)}
\end{equation}

\begin{figure}[]
\centering
\includegraphics[width=3in]{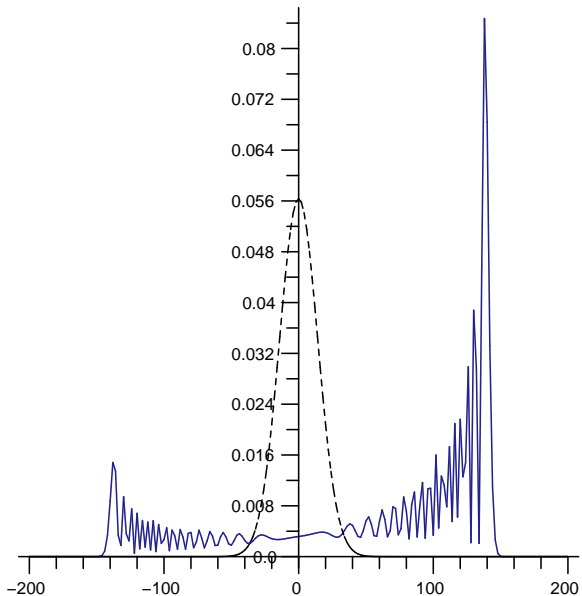}
\caption{Overlay of distributions of classical (dotted black line) and quantum (solid blue line) random walks running for 200 time steps.}\label{fig:qw_overlay}
\end{figure}

A celebrated result \cite{nayak-2000,ambainis2001} is that this homogeneous quantum walk spreads linearly in time, quadratically faster than classical walks. The shape of the distribution is also very different, as can be seen in Figure \ref{fig:qw_overlay}.

\section{\label{sec:section2}Inhomogeneous Quantum Walks}

Inhomogeneous quantum walks differ from the ones described above in that we allow the coin operator to depend on the position register and the coin register, rather than only on the coin register of the state space. This leads to many interesting new behaviours, such as walks that remain bounded in a certain region for all time, as we shall see.

We will define an inhomogeneous quantum walk in a similar manner to the standard quantum walk, however we now allow the coin operator $C$ to be dependent on $m$, the current position of the walk:
\begin{equation}
W = S \left( \displaystyle \sum_m \ket{m}\bra{m} \otimes C_m \right)
\end{equation}

In the case of walks on the line $C_m$ could be an arbitrary unitary operator on the two-dimensional coin space.  Indeed, there
is no need to restrict to walks in which one only moves to the nearest neighbours: more generally,
one could allow there to be transitions from a given point to any other point on the line.  However for the purposes
of most of our discussion of walks on the line we will focus on the simplest case of a transitions to nearest neighbours.

As a first step in this process we will consider a family of examples in which the coin operator is periodic, with the coin operator whose
matrix in the basis $\{\ket L,\ket R \}$ given by:
\begin{equation}
\label{eqn-cos_sin_matrix}
 \left( \begin{array}{cc} \cos{ \left( \frac{n \pi}{k} \right) } & -\sin{ \left( \frac{n \pi}{k} \right) } \\ \sin{ \left( \frac{n \pi}{k} \right) } & \cos{ \left( \frac{n \pi}{k} \right) } \end{array} \right)
\end{equation}
Here $k $ is an arbitrary positive integer; the period of this walk is $2k$. More explicitly, this means that this coin operator transforms states in the following way:
\begin{align}
C_n \ket{n,L} & = \cos{\left(\frac{n\pi}{k}\right)} \ket{n,L} + \sin{\left(\frac{n\pi}{k}\right)} \ket{n,R} \\
C_n \ket{n,R} & = -\sin{\left(\frac{n\pi}{k}\right)} \ket{n,L} + \cos{\left(\frac{n\pi}{k}\right)} \ket{n,R}
\end{align}

\begin{figure}[]
\centering
\includegraphics[width=2.5in, height=2in]{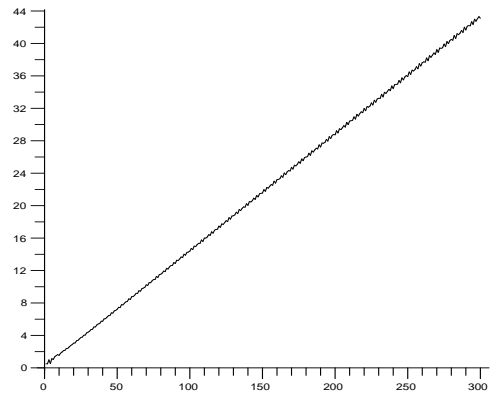}
\caption{Standard deviation of quantum walk with coin from equation (\ref{eqn-cos_sin_matrix}), initial state $\frac{1}{\sqrt{2}}(\ket{0,L}+\ket{0,R})$ and $k=3$ against time steps.}\label{fig:qw_sd_npi3_300}
\end{figure}

\begin{figure}[!htb]
\centering
\includegraphics[keepaspectratio=false, width=3in, height=2in]{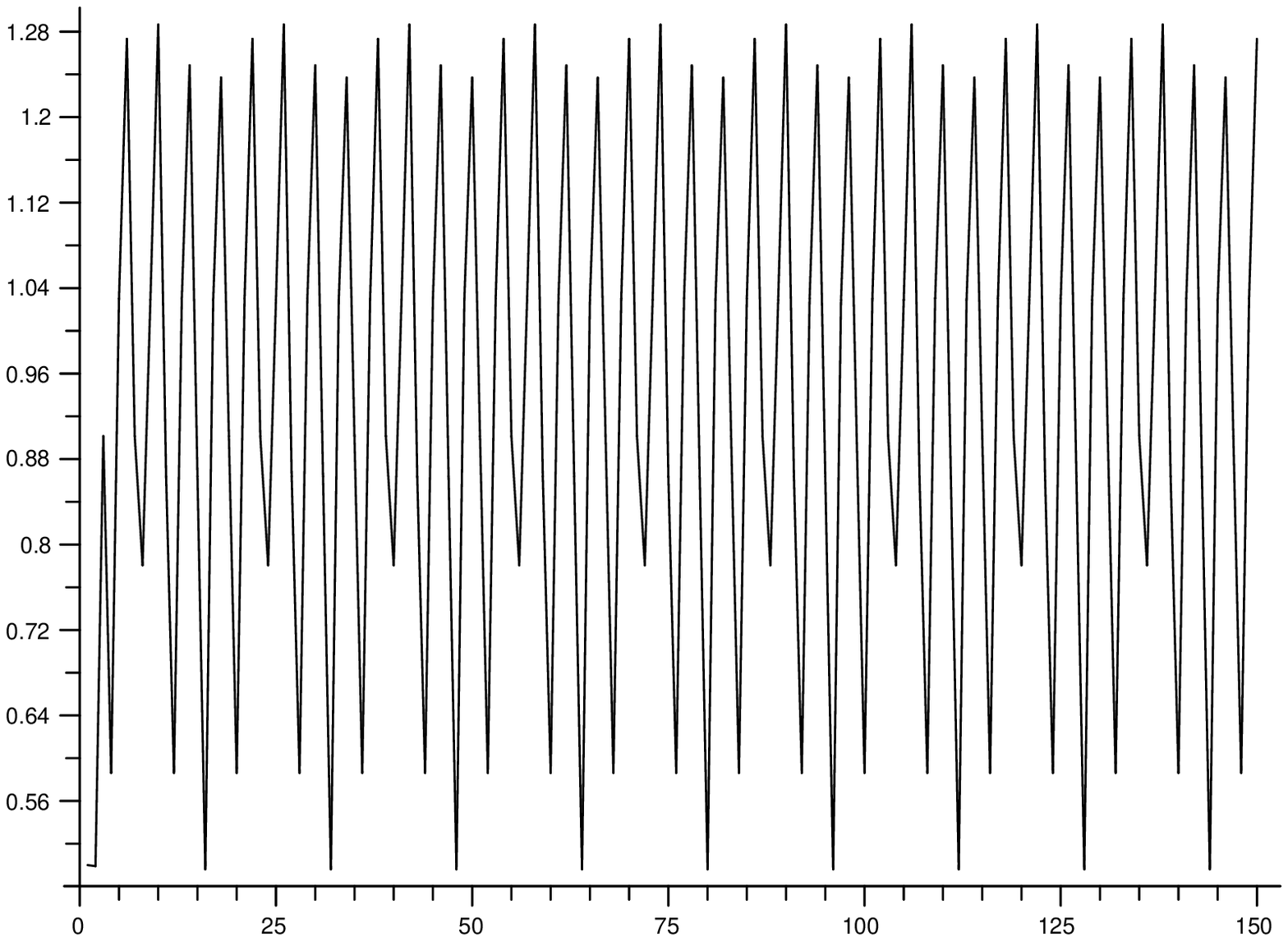}
\caption{Standard deviation of quantum walk with coin from equation (\ref{eqn-cos_sin_matrix}), initial state $\frac{1}{\sqrt{2}}(\ket{0,L}+\ket{0,R})$ and $k=4$ against time steps.}\label{fig:qw_sd_npi4_150}
\end{figure}

\begin{figure}[!htb]
\centering
\includegraphics[keepaspectratio=false, width=3in, height=2in]{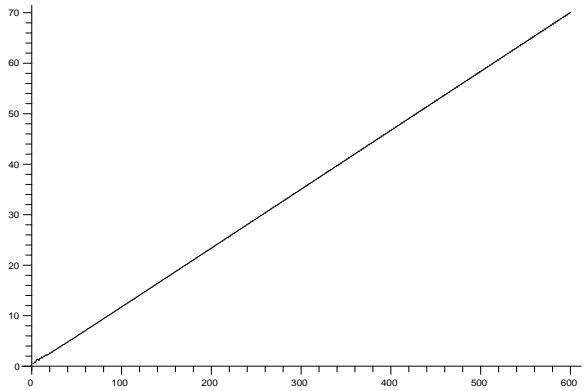}
\caption{Standard deviation of quantum walk with coin from equation (\ref{eqn-cos_sin_matrix}), initial state $\frac{1}{\sqrt{2}}(\ket{0,L}+\ket{0,R})$ and $k=5$ against time steps.}\label{fig:qw_sd_npi5_600}
\end{figure}

Plots of the standard deviation against the number of time steps for the coin in equation (\ref{eqn-cos_sin_matrix}) can be seen  in figures \ref{fig:qw_sd_npi3_300}, \ref{fig:qw_sd_npi4_150} and \ref{fig:qw_sd_npi5_600}. They provide some insight into the behaviour that we might expect from these walks, figures \ref{fig:qw_sd_npi3_300} and \ref{fig:qw_sd_npi5_600} appear to be moving linearly in time and figure \ref{fig:qw_sd_npi4_150} appears to be periodic in time. In this section we analyze these cases in detail.

Firstly we consider the general conditions for a walk to be restricted to a finite region of the line for all time:
\newline
\textbf{Lemma 2.1}
\textit{Let the initial state of the walk to be of the form
\begin{equation}
\label{simple-initial-state}
\ket {n_0} \left( a_0\ket L + b_0 \ket R\right);
\end{equation}
then an inhomogeneous quantum random walk with a unitary coin is bounded iff there exists to the left and the right of $n_0$ a coin with matrix of the form }
\begin{equation}
\left( \begin{array}{ccc}
0 & e^{i\theta} \\
e^{i\phi} & 0  \end{array} \right)
\end{equation}
\newline
\textit{Proof: }
First assume that the walk is bounded. Then there exists a least upper bound and a greatest lower bound for the walk. Let $n_l$ denote the position of the greatest lower bound and $n_u$ denote the position of the least upper bound. Since $n_l$ is the greatest lower bound for the walk, the walk must reach this point and there is zero probability of moving further left. Hence the coin at $n_l$ must be of the form:
\begin{equation}
C_{n_l} = \left( \begin{array}{ccc}
0 & a \\
b & c  \end{array} \right)
\end{equation}
Since the coin is unitary, it must mean that the bottom right component is zero as well. Thus the matrix must be of the following form on the lower bound:
\begin{equation}
\label{lower-coin}
C_{n_l} = \left( \begin{array}{ccc}
0 & e^{i\theta} \\
e^{i\phi} & 0  \end{array} \right)
\end{equation}
Analogously, since $n_u$ is the least upper bound for the walk, it must mean that at this point there is zero probability of moving to the right. It follows that the coin at $n_u$ must be of the form:
\begin{equation}
C_{n_u} = \left( \begin{array}{ccc}
e & f \\
g & 0  \end{array} \right)
\end{equation}
Using the same argument as for the lower bound, the fact that the coin is unitary means that the coin at the upper bound must have the following form:
\begin{equation}
\label{upper-coin}
C_{n_u} = \left( \begin{array}{ccc}
0 & e^{i\theta'} \\
e^{i\phi'} & 0  \end{array} \right)
\end{equation}
Therefore, if a random walk is bounded, it must have coins of the above form both above and below the initial state.

Conversely, we now assume that there exists a coin with a matrix of the  form (\ref{lower-coin}) both to the left and the right of the initial state. Firstly  consider the coin with form (\ref{lower-coin}) to the right of the initial state. The first time the walk approaches this position on the line, it must come from the left. However, the coin above makes everything that is travelling to the right change direction and travel to the left. Thus if amplitude approaches from the left, it will never continue on to the right past that point. Hence this must be an upper bound for the walk.

Similarly, if we consider the coin with the above form to the left of the initial state, it must first be approached from the right. Since the form of the above matrix flips the direction of movement, everything approaching from the right will be reflected. This means that this position on the line must be a lower bound for the walk. Hence, the quantum walk is bounded.$\Box$

Note that a general initial state is a superposition of states of the type (\ref{simple-initial-state}); then the condition for boundedness applies separately
to each term in the superposition (so, in general, one could have parts of the wave-function that remain bounded in a region and other parts that escape).

As an example, applying this lemma to the coin we defined in equation (\ref{eqn-cos_sin_matrix}), we can see that the walk will be bounded for even $k$ and unbounded for odd $k$.

Now that we understand when a walk will be bounded, we can move on to explore the behaviour when it is unbounded. We would like to calculate the standard deviation of a general periodic walk in order to work out how fast it spreads. We are particularly interested in the family of walks with coin matrix (\ref{eqn-cos_sin_matrix}).  These have even period for any integer $k$.   Initially we set up the problem for general walks with even period $2\Delta$.  Thus we define a general periodic unitary coin $C_{n,2\Delta}$ with period $2\Delta,\ \Delta\in \mathbb{Z}$ as follows:
\begin{equation}
\label{eqn-gen_periodic_coin}
C_{n, \Delta} = \left( \begin{array}{cc} \alpha_{n,\Delta} & -e^{i\theta_n}\bar{\beta}_{n,\Delta} \\ \beta_{n,\Delta} & e^{i\theta_n}\bar{\alpha}_{n,\Delta} \end{array} \right)
\end{equation}
Where $C_{n+2\Delta,\Delta} = C_{n,\Delta}$ and $|\alpha_{n,\Delta}|^2 + |\beta_{n,\Delta}|^2 = 1$ for all $n$. Henceforth we shall omit the $\Delta$ in the labelling of $C$, $\alpha$ and $\beta$ in order to keep the notation tidy, i.e $C_{n,\Delta} = C_n$.

\subsection{The Period 6 Case}
\label{sec:sec2period6}

In order to illustrate the method clearly, we will first use the technique for the case where the period is $6$, i.e. where $\Delta=3$. We will then move on to define the general case in section \ref{sec:sec2general}.

Starting with the definition of the coin in equation (\ref{eqn-gen_periodic_coin}) with $2\Delta=6$, one can derive a recurrence relation for the wave function:
\begin{equation}
\label{eqn_recurrence_relation}
\Psi(n,t+1) = C^+_{n-1}\Psi(n-1,t) + C^-_{n+1}\Psi(n+1,t)
\end{equation}
Where:
\begin{align}
C^+_n = \left( \begin{array}{cc} 0 & 0 \\ \beta_{n} & e^{i\theta_n}\bar{\alpha}_n \end{array} \right) & , C^-_n = \left( \begin{array}{cc} \alpha_n & -e^{i\theta_n}\bar{\beta}_n \\ 0 & 0 \end{array} \right) \\
\Psi(n,t) & = \left( \begin{array}{c} \psi_L(n,t) \\ \psi_R(n,t) \end{array} \right)
\end{align}
Here $\psi_L(n,t)$ is the amplitude of the wave function travelling left at position $n$ and time $t$, and $\psi_R(n,t)$ is the corresponding amplitude travelling right. i.e.
\begin{equation}
\ket{\psi(t)}=\sum_{n=-\infty}^\infty \big(\psi_L(n,t)\ket{n,L}+ \psi_R(n,t)\ket{n,R}\big)
\end{equation}

\begin{figure}[t]
\centering
\includegraphics[width=3in, height=2in]{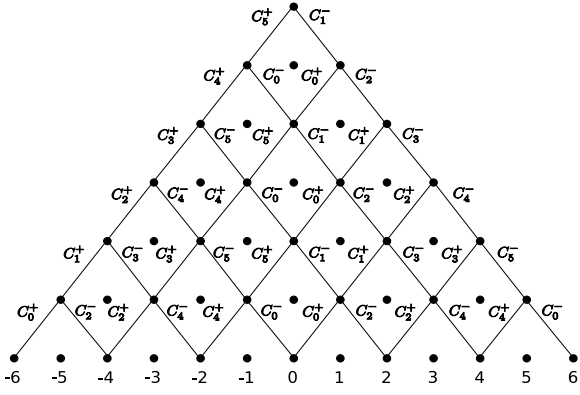}
\caption{Graphical representation of the computation of the coefficients $c_{0,j}$ in equation (\ref{eqn_sub_recur_relation}) }\label{c-i-j}
\end{figure}

Recursively substituting the recurrence equation into itself six times, starting from $\Psi(6n,t+6)$, $\Psi(6n+2,t+6)$ and $\Psi(6n+4,t+6)$ yields equations of the form:
\begin{equation}
\begin{split}
\label{eqn_sub_recur_relation}
\Psi (6n, t+6) & = \sum_{j=0}^{6} c_{0,j} \Psi(6(n-1)+2j, t) \\
\Psi (6n+2, t+6) & = \sum_{j=0}^{6} c_{1,j} \Psi(6(n-1)+2j+2, t) \\
\Psi (6n+4, t+6) & = \sum_{j=0}^{6} c_{2,j} \Psi(6(n-1)+2j+4, t)
\end{split}
\end{equation}
where the $c_{i,j}$ are constant (i.e. independent of $n$) $2\times2$ matrices given by sums of products of $C_p^+$ and $C_q^-$.  There is a simple way of computing these coefficients by referring to Figure \ref{c-i-j} . For example
\begin{eqnarray}
c_{0,0}&=& C^+_5 C^+_4 C^+_3 C^+_2 C^+_1 C^+_0\nonumber\\
c_{0,1}&=& C^+_5 C^+_4 C^+_3 C^+_2 C^+_1 C^-_2\nonumber\\
& & + C^+_5 C^+_4 C^+_3 C^+_2 C^-_3 C^+_2\nonumber\\
& & + C^+_5 C^+_4 C^+_3 C^-_4 C^+_3 C^+_2\nonumber\\
& & + C^+_5 C^+_4 C^-_5 C^+_4 C^+_3 C^+_2\nonumber\\
& & + C^+_5 C^-_0 C^+_5 C^+_4 C^+_3 C^+_2\nonumber\\
& & + C^-_1 C^+_0 C^+_5 C^+_4 C^+_3 C^+_2
\end{eqnarray}

We will change notation at this point to make the equations more concise. We define $\Psi(6n+2i,t) := \Psi_i(n,t)$ and grouping certain terms in the sum allows us to write the previous equation as follows:
\begin{equation}
\begin{split}
\label{eqn_sub_recur_relation_0_newnot}
\Psi_0(n, t+6) & = c_{0,0}\Psi_0(n-1, t) + c_{0,1}\Psi_1(n-1, t) \\
 & + c_{0,2}\Psi_2(n-1, t) + c_{0,3}\Psi_0(n, t) \\
 & + c_{0,4}\Psi_1(n, t) + c_{0,5}\Psi_2(n, t) \\
 & + c_{0,6}\Psi_0(n+1, t)
\end{split}
\end{equation}

\begin{equation}
\begin{split}
\label{eqn_sub_recur_relation_1_newnot}
\Psi_1(n, t+6) & = c_{1,0}\Psi_1(n-1, t) + c_{1,1}\Psi_2(n-1, t) \\
 & + c_{1,2}\Psi_0(n, t) + c_{1,3}\Psi_1(n, t) \\
 & + c_{1,4}\Psi_2(n, t) + c_{1,5}\Psi_0(n+1, t) \\
 & + c_{1,6}\Psi_1(n+1, t)
\end{split}
\end{equation}

\begin{equation}
\begin{split}
\label{eqn_sub_recur_relation_2_newnot}
\Psi_2(n, t+6) & = c_{2,0}\Psi_2(n-1, t) + c_{2,1}\Psi_0(n, t) \\
 & + c_{2,2}\Psi_1(n, t) + c_{2,3}\Psi_2(n, t) \\
 & + c_{2,4}\Psi_0(n+1, t) + c_{2,5}\Psi_1(n+1, t) \\
 & + c_{2,6}\Psi_2(n+1, t) \\
 &
\end{split}
\end{equation}
Discrete Fourier transforming these equations yields the following:
\begin{equation}
\begin{split}
\tilde{\Psi}_0(\omega, t+6) & = \left(c_{0,0}e^{-i\omega}+c_{0,3}+c_{0,6}e^{i\omega}\right) \tilde{\Psi}_0(\omega,t) \\
 & + \left(c_{0,1}e^{-i\omega}+c_{0,4}\right) \tilde{\Psi}_1(\omega,t) \\
 & + \left(c_{0,2}e^{-i\omega}+c_{0,5}\right) \tilde{\Psi}_2(\omega,t)
\end{split}
\end{equation}
\begin{equation}
\begin{split}
\tilde{\Psi}_1(\omega, t+6) & = \left(c_{1,2}+c_{1,5}e^{i\omega}\right) \tilde{\Psi}_0(\omega,t) \\
 & + \left(c_{1,0}e^{-i\omega}+c_{1,3}+c_{1,6}e^{i\omega}\right) \tilde{\Psi}_1(\omega,t) \\
 & + \left(c_{1,1}e^{-i\omega}+c_{1,4}\right) \tilde{\Psi}_2(\omega,t)
\end{split}
\end{equation}
\begin{equation}
\begin{split}
\tilde{\Psi}_2(\omega, t+6) & = \left(c_{2,1}+c_{2,4}e^{i\omega}\right) \tilde{\Psi}_0(\omega,t) \\
 & + \left(c_{2,2}+c_{2,5}e^{i\omega}\right) \tilde{\Psi}_1(\omega,t) \\
 & + \left(c_{2,0}e^{-i\omega}+c_{2,3}+c_{2,6}e^{i\omega}\right) \tilde{\Psi}_2(\omega,t),
\end{split}
\end{equation}
where
\begin{eqnarray}
\tilde f(\omega)&=& \sum_{n=-\infty}^\infty e^{-in\omega}f(n)\nonumber\\
f(n)&=&\frac{1}{2\pi}\int_{-\pi}^\pi d\omega\ e^{in\omega} \tilde f(\omega).
\end{eqnarray}

Thus we can give an expression for $\tilde{\Psi} (\omega,6T+\gamma)$ in Fourier space where $\{ \gamma =0,1,2,3,4,5 \}$:
\begin{equation}
\label{eqn-U_matrix}
\left( \begin{array}{ccc} \tilde{\Psi}_0 (\omega, 6T+\gamma ) \\ \tilde{\Psi}_1 (\omega, 6T+\gamma ) \\ \tilde{\Psi}_2 (\omega, 6T+\gamma ) \\\end{array} \right) = U^T
\left( \begin{array}{ccc} \tilde{\Psi}_0 (\omega, \gamma ) \\ \tilde{\Psi}_1 (\omega, \gamma ) \\ \tilde{\Psi}_2 (\omega, \gamma ) \\\end{array} \right)
\end{equation}
\begin{widetext}
Where the $6\times 6$ matrix $U$ is:
\begin{equation}
U = \left( \begin{array}{ccc} c_{0,0}e^{-i\omega}+c_{0,3}+c_{0,6}e^{i\omega} & c_{0,1}e^{-i\omega}+c_{0,4} & c_{0,2}e^{-i\omega}+c_{0,5} \\
 c_{1,2}+c_{1,5}e^{i\omega} & c_{1,0}e^{-i\omega}+c_{1,3}+c_{1,6}e^{i\omega} & c_{1,1}e^{-i\omega}+c_{1,4} \\
 c_{2,1}+c_{2,4}e^{i\omega} & c_{2,2}+c_{2,5}e^{i\omega} & c_{2,0}e^{-i\omega}+c_{2,3}+c_{2,6}e^{i\omega} \end{array} \right)
\end{equation}
\end{widetext}

The matrix $U$ in equation (\ref{eqn-U_matrix}), is unitary, which allows us to write it as:
\begin{equation}
U(\omega) = U_0(\omega) D(\omega) U_0^\dagger(\omega)
\end{equation}
where $U_0$ is a unitary matrix composed of the eigenvectors of $U$ and $D$ is a diagonal matrix containing the eigenvalues of $U$. This allows us to rewrite $U^T$ in the following way:
\begin{equation}
U^T(\omega) = U_0(\omega) D^T(\omega) U_0^\dagger(\omega).
\end{equation}
Since the eigenvalues of a unitary matrix lie on the unit circle on the complex plane,
\begin{equation}
(D^T)_{jk} = \delta_{jk} e^{i\lambda_j(\omega)T},
\end{equation}
where $\delta_{jk}$ is the Kronecker delta.

This yields integral expressions for the wave function at arbitrary time and position:
\begin{equation}
\label{eqn-fourier_integral_L0}
\psi_{0L} (n, 6T) = \frac{1}{2 \pi} \int_{-\pi}^\pi \sum_{l=1}^6 \alpha_{1,l}(\omega)e^{iT(\lambda_l(\omega)+\omega \beta)} d\omega,
\end{equation}
\begin{equation}
\label{eqn-fourier_integral_R0}
\psi_{0R} (n, 6T) = \frac{1}{2 \pi} \int_{-\pi}^\pi \sum_{l=1}^6 \alpha_{2,l}(\omega)e^{iT(\lambda_l(\omega)+\omega \beta)} d\omega,
\end{equation}
\begin{equation}
\label{eqn-fourier_integral_L1}
\psi_{1L} (n, 6T) = \frac{1}{2 \pi} \int_{-\pi}^\pi \sum_{l=1}^6 \alpha_{3,l}(\omega)e^{iT(\lambda_l(\omega)+\omega \beta)} d\omega,
\end{equation}
\begin{equation}
\label{eqn-fourier_integral_R1}
\psi_{1R} (n, 6T) = \frac{1}{2 \pi} \int_{-\pi}^\pi \sum_{l=1}^6 \alpha_{4,l}(\omega)e^{iT(\lambda_l(\omega)+\omega \beta)} d\omega,
\end{equation}
\begin{equation}
\label{eqn-fourier_integral_L2}
\psi_{2L} (n, 6T) = \frac{1}{2 \pi} \int_{-\pi}^\pi \sum_{l=1}^6 \alpha_{5,l}(\omega)e^{iT(\lambda_l(\omega)+\omega \beta)} d\omega,
\end{equation}
\begin{equation}
\label{eqn-fourier_integral_R2}
\psi_{2R} (n, 6T) = \frac{1}{2 \pi} \int_{-\pi}^\pi \sum_{l=1}^6 \alpha_{6,l}(\omega)e^{iT(\lambda_l(\omega)+\omega \beta)} d\omega,
\end{equation}
where:
\begin{equation}
\begin{split}
&\alpha_{ij}(\omega)  = \\ &(U_0)_{ij} \left[ \sum_{m=0}^2 (U_0^\dagger)_{j,2m+1} \tilde{\psi}_{lL}(\omega, 0)  + (U_0^\dagger)_{j,2m+2} \tilde{\psi}_{lR}(\omega, 0) \right],
\end{split}
\end{equation}
and
\begin{equation}
\beta=n/T.
\end{equation}

We now use the integral expressions in equations (\ref{eqn-fourier_integral_L0}) to (\ref{eqn-fourier_integral_R2}) to calculate the first and second moments of the probability distribution on the line. From these we can calculate the standard deviation and thus the rate at which the walk spreads. It is important to note here that we compute these moments  only  for a number of time steps that is a multiple of six.

The probability $p$ that the walk is at position $6n+2j$ after $6T$ time steps is given by:
\begin{equation}
p(6n+2j, 6T) = |\psi_{jL}(n,6T)|^2 + |\psi_{jR}(n,6T)|^2
\end{equation}
Hence we can calculate the first moment after $6T$ time steps:
\begin{equation}
\mathbb{E}(N,6T) = \sum_{n=-\infty}^\infty \sum_{j=0}^2 (6n+2j) p(6n+2j, 6T).
\end{equation}
Or more explicity,
\begin{widetext}
\begin{equation}
\label{eqn:expect_N}
\mathbb{E}(N,6T) = \sum_{n=-\infty}^\infty \sum_{j=1}^6 \sum_{l_1,l_2=1}^6 \frac{6n+2\lfloor \frac{j-1}{2} \rfloor}{(2\pi)^2} \int_{-\pi}^{\pi} d\omega_1 \int_{-\pi}^{\pi} d\omega_2 e^{i(\omega_1 - \omega_2)n} \alpha_{j,l_1}(\omega_1)\alpha_{j,l_2}^*(\omega_2) e^{iT(\lambda_{l_1}(\omega_1) - \lambda_{l_2}(\omega_2))}
\end{equation}
\end{widetext}
Where $\lfloor \frac{j-1}{2} \rfloor$ is the truncated integer part of $\frac{j-1}{2}$.

We are interested in the leading order of equation (\ref{eqn:expect_N}) in $T$ as $T \rightarrow \infty$ because we are calculating the long time variance of the system. We can simplify this expression by noting that the integrals with coefficient $\frac{2}{(2\pi)^2}\lfloor\frac{j-1}{2}\rfloor$ in equation (\ref{eqn:expect_N}) are of order $O({1})$ using the method of stationary phase.

As we will show that $O(1)$ is not the leading order in $T$, we can discard it from our calculation. We can further simplify the expression using:
\begin{equation}
\frac{1}{2\pi} \displaystyle \sum_{x=-\infty}^\infty x^n e^{i(\omega - \omega')x} = (-i)^n \delta^{(n)}(\omega-\omega')
\end{equation}
This yields a much simpler equation for the dominant terms in the expectation of $N$:
\begin{equation}
\begin{split}
\mathbb{E}(N,6T) & \simeq \displaystyle \frac{3i}{\pi} \left[ \int_{-\pi}^{\pi} d\omega_1 \sum_{j,l_1,l_2=1}^6 \left( iT\lambda^\prime_{l_1}(\omega_1)\alpha_{jl_1}(\omega_1) \right.\right. \\ & \left.\left. + \alpha_{jl_1}'(\omega_1)  \right) \alpha_{jl_2}^* e^{iT(\lambda_{l_1}(\omega_1) - \lambda_{l_2}(\omega_1))} \right]
\end{split}
\end{equation}

For $l_1=l_2$, the exponentials disappear and we are left with terms linear in $T$ or constant after the integral over $\omega_1$ has been performed. Hence these terms give a leading order term $\mathbb{E}(N,6T) \simeq d_0T$  assuming that $d_0$ does not vanish. When $l_1 \neq l_2$ we can apply the method of stationary phase as $T \rightarrow \infty$,  to each term, assuming that the original $U$ (\ref{eqn-U_matrix}) is non-degenerate (as we will find in the example of particular interest to us - see below) and find that each term is, at most, $O(\sqrt{T})$. Hence we can say that the expectation of $N$ has leading order
\begin{equation}
\mathbb{E}(N,6T)\simeq -6T\int_{-\pi}^\pi \frac{d\omega}{2 \pi}\sum_{j,l=1}^6 |\alpha_{jl}(\omega)|^2\lambda_l^\prime(\omega).
\end{equation}
as $T\rightarrow \infty$.

The second moment for an even number of time steps can be calculated in a similar way using:
\begin{equation}
\mathbb{E}(N^2,6T) = \sum_{n}\sum_{j=0}^2 (6n+2j)^2 p(6n+2j, 6T)
\end{equation}

However, the $24nj$ and $4j^2$ coefficients of $p$ are at most $O(T)$ and $O(1)$ respectively. Since we will show that neither of these are leading order in $T$ we can ignore them from the calculation and end up with an expression for the leading order behaviour for the second moment:
\begin{equation}
\begin{split}
&\mathbb{E}(N^2,6T)\\
 &\quad \simeq \frac{36}{2\pi} \int_{-\pi}^{\pi} d\omega \sum_{j,l_1,l_2=1}^6 \left( T^2\lambda_{l_1}'^2\alpha_{jl_1} \right. \\ & \quad - iT\left( \lambda_{l_1}''\alpha_{jl_1} + 2\lambda_{l_1}'\alpha_{jl_1}' \right) - \left. \alpha_{jl_1}'' \right) \alpha_{jl_2}^* e^{iT(\lambda_{l_1} - \lambda_{l_2})}
\end{split}
\end{equation}

Using the same reasoning as before, when $l_1=l_2$ we have terms like $d_2T^2 + d_3T + d_4$ where $d_2$,$d_3$ and $d_4$ are  constants. Using the method of stationary phase for the other terms yields terms of order less than or equal to $T^{\frac{3}{2}}$. Hence, assuming $d_2$ does not vanish and  the leading order term in $\mathbb{E}(N^2,6T) $ behaves like $T^2$ as $T\rightarrow\infty$.

Collecting the leading order terms together we find that the leading order term in the variance is
\begin{equation}
\label{Var}
{\rm Var}(N,6T)\simeq \frac{36 T^2}{2 \pi}\int_{-\pi}^\pi d\omega\sum_{j,l=1}^6 |\alpha_{jl}(\omega)|^2\big(\lambda_l^\prime(\omega)-\mu\big)^2,
\end{equation}
where
\begin{equation}
\mu= \frac{1}{2 \pi}\int_{-\pi}^\pi d\omega\sum_{j,l=1}^6 |\alpha_{jl}(\omega)|^2\lambda_l^\prime(\omega).
\end{equation}

The expression (\ref{Var}) is valid for any coin of period six.  And given its form - an integral of a sum of terms, each of which is a product of two positive terms - one can see why one might expect, typically, that the leading order behaviour of the standard deviation will
indeed to be linear in $T$.  We have not been able to get closed form analytic expressions for $\alpha_{jl}(\omega)$ or $\lambda_l^\prime(\omega)$ (they
involve diagonalising a $6\times 6$ matrix).
However in the case that the coin $C_n$ has the matrix with the simple form (with period 6)
\begin{equation}
\left( \begin{array}{cc} \cos{ \left( \frac{n \pi}{3} \right) } & -\sin{ \left( \frac{n \pi}{3} \right) } \\ \sin{ \left( \frac{n \pi}{3} \right) } & \cos{ \left( \frac{n \pi}{3} \right) } \end{array} \right),
\end{equation}
and where the initial state is
\begin{equation}
\ket{\psi(0)}=\ket 0 \left(\frac{\ket L + \ket R}{\sqrt 2} \right),
\end{equation}
we give figures below plotting $\sum_{j=1}^6|\alpha_{jl}(\omega)|^2$ and $\lambda_l^\prime(\omega)$ for $l=1...6$.  These show that the variance does indeed
have leading order behaviour proportional to $T^2$, and hence the spread of the walk is proportional to $T$, as observed in the earlier figure \ref{fig:qw_sd_npi3_300}.
\begin{figure}[htb]
\centering
\includegraphics[keepaspectratio=false, width=3in, height=2in]{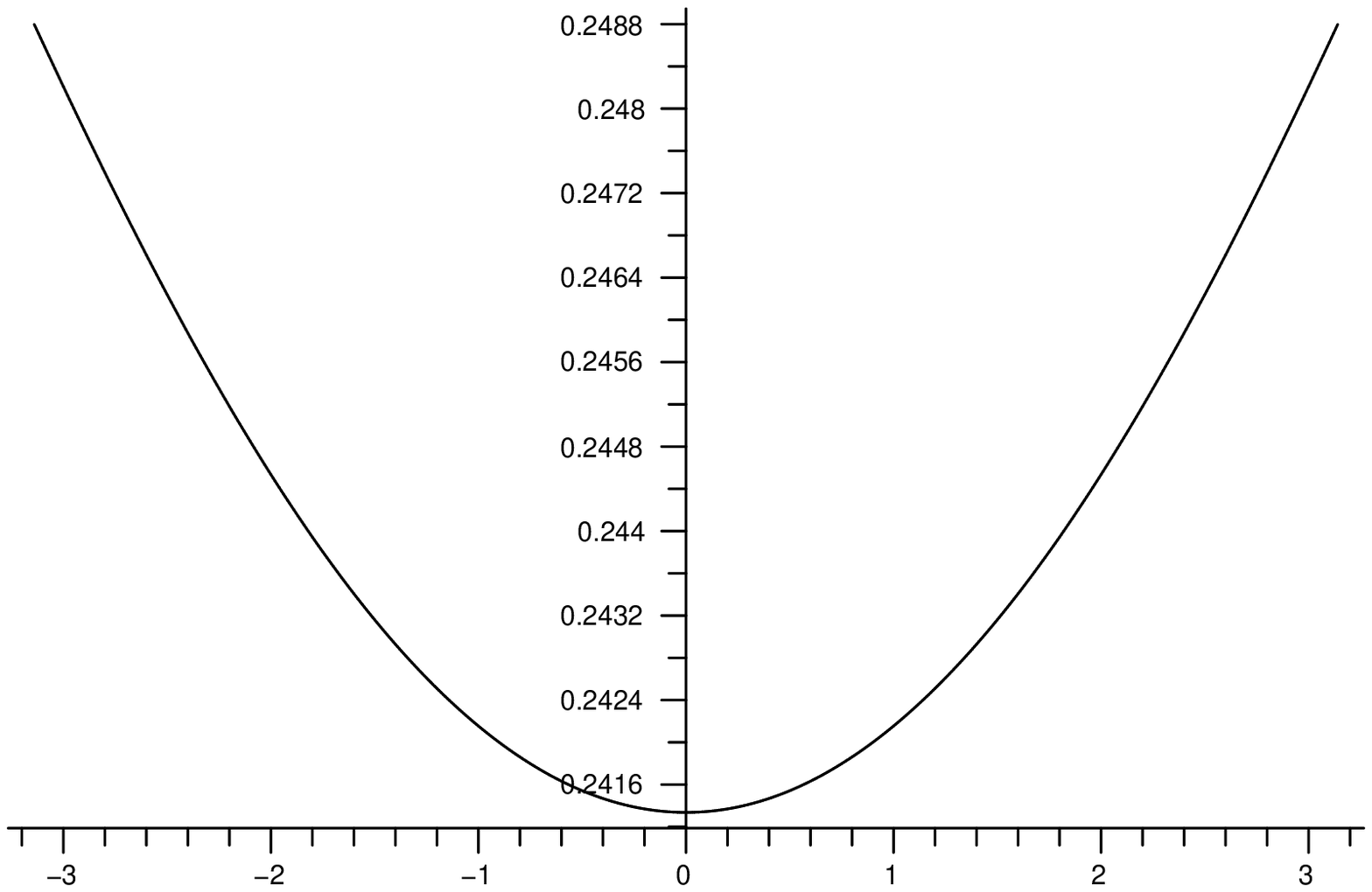}
\caption{Plot of $\sum_{j=1}^6|\alpha_{j1}(\omega)|^2$ and $\sum_{j=1}^6|\alpha_{j2}(\omega)|^2$. Graphs of both are identical, hence the single graph representing either sum. The initial state is $\frac{1}{\sqrt{2}}\left(\ket{0,L} + \ket{0,R}\right)$ }\label{fig:alpha12}
\end{figure}

\begin{figure}[htb]
\centering
\includegraphics[keepaspectratio=false, width=3in, height=2in]{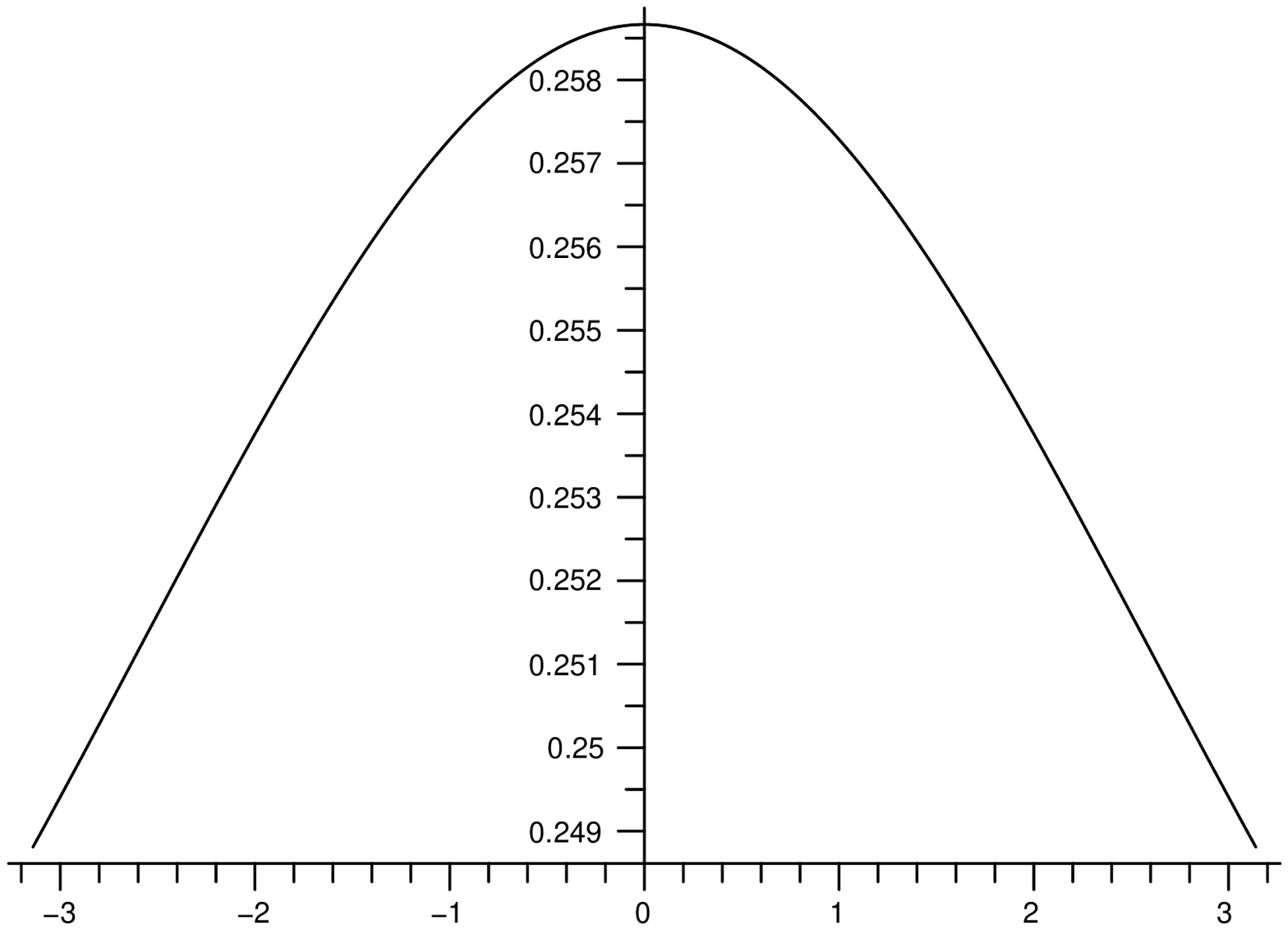}
\caption{Plot of $\sum_{j=1}^6|\alpha_{j3}(\omega)|^2$ and $\sum_{j=1}^6|\alpha_{j4}(\omega)|^2$. Graphs of both are identical, hence the single graph representing either sum. The initial state is $\frac{1}{\sqrt{2}}\left(\ket{0,L} + \ket{0,R}\right)$ }\label{fig:alpha34}
\end{figure}

\begin{figure}[htb]
\centering
\includegraphics[keepaspectratio=false, width=3in, height=2in]{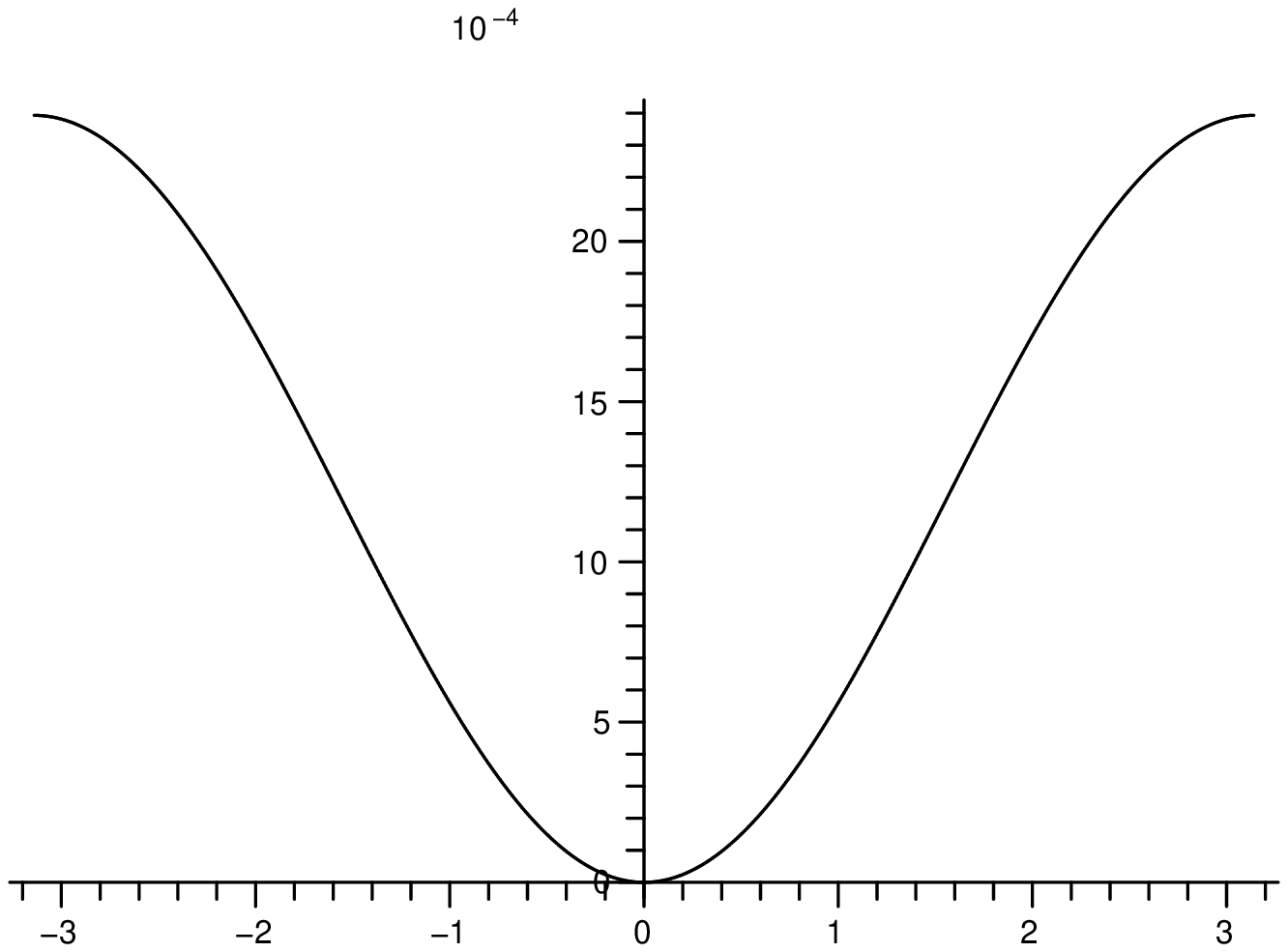}
\caption{Plot of $\sum_{j=1}^6|\alpha_{j5}(\omega)|^2$ and $\sum_{j=1}^6|\alpha_{j6}(\omega)|^2$. Graphs of both are identical, hence the single graph representing either sum. The initial state is $\frac{1}{\sqrt{2}}\left(\ket{0,L} + \ket{0,R}\right)$ }\label{fig:alpha56}
\end{figure}
\begin{figure}[htb]
\centering
\includegraphics[keepaspectratio=false, width=3in, height=2in]{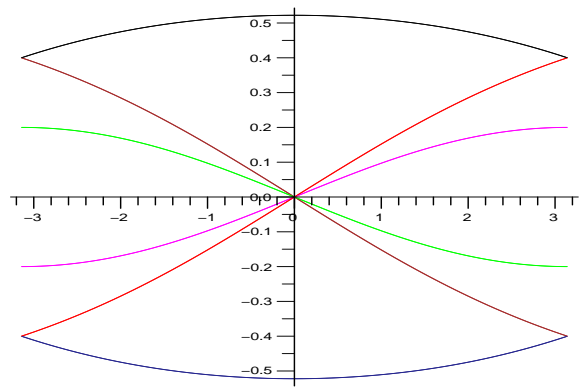}
\caption{Graphs of $\lambda_l^\prime(\omega)$. Each line represents the graph of a different value of $l$. The black line is $\lambda_\prime^\prime$, the blue line is $\lambda_2^\prime$, the brown line is $\lambda_3^\prime$, the red line is $\lambda_4^\prime$, the green line is $\lambda_5^\prime$ and the magenta line is $\lambda_6^\prime$.}\label{fig:lambda-primed}
\end{figure}

\subsection{The General Case}
\label{sec:sec2general}

Very similar calculations can be performed for the period $2\Delta$ case, for general $\Delta$, leading to an expression for the leading order behaviour exactly analogous
to that for the period six case:
\begin{equation}
\begin{split}
\label{VarK}
{\rm Var} & (N,2\Delta T) \simeq \\ & \frac{(2\Delta)^2 T^2}{2 \pi}\int_{-\pi}^\pi d\omega\sum_{j,l=1}^{2\Delta} |\alpha_{jl}(\omega)|^2\big(\lambda_l^\prime(\omega)-\mu\big)^2,
\end{split}
\end{equation}
where
\begin{equation}
\mu= \frac{1}{2 \pi}\int_{-\pi}^\pi d\omega\sum_{j,l=1}^{2\Delta} |\alpha_{jl}(\omega)|^2\lambda_l^\prime(\omega).
\end{equation}

The functions $\alpha_{jl}(\omega)$ and  $\lambda_l(\omega)$ arise, as in the period six case, from the initial condition and diagonalization of the
$2\Delta\times 2\Delta$ unitary matrix defining the time evolution:
\begin{equation}
\label{eqn-U_matrix_general}
\left( \begin{array}{ccc} \tilde{\Psi}_0 (\omega, 2\Delta T+\gamma ) \\ \vdots \\ \tilde{\Psi}_{\Delta-1} (\omega, 2\Delta T+\gamma ) \\\end{array} \right) = U^T
\left( \begin{array}{ccc} \tilde{\Psi}_0 (\omega, \gamma ) \\ \vdots \\ \tilde{\Psi}_{\Delta-1} (\omega, \gamma ) \\\end{array} \right)
\end{equation}

Earlier it was shown  that walks with coins of the form given in equation (\ref{eqn-cos_sin_matrix}) with $k$ even are bounded for all time (see also Fig. \ref{fig:qw_sd_npi4_150}). A natural question that arises is how  this fits in to the derivation above.

Obviously the \lq\lq leading term\rq\rq\ in (\ref{VarK}) must be zero.  Clearly there are various ways that this could occur; in particular (\ref{VarK})
will be zero if the $\lambda_l(\omega)$ are in fact independent of $\omega$.  By direct calculation we have checked that this is indeed the case for
$k=2$ and $k=4$.

\section{\label{sec:section3} Quantum Walks Defined Via Double-Reflection}

In \cite{szegedy}, Szegedy proposed a  method for defining the quantum walk on a graph starting with a classical Markov chain. The work was generalized in \cite{magniez2007a}.  For reasons that will be clear shortly, we refer to these quantum walks as \lq\lq double-reflection\rq\rq\ walks.

We now compare the construction we have been using with that in \cite{szegedy,magniez2007a} for the particular case of motion on the line.  The state space used in \cite{szegedy,magniez2007a}
consists of two copies of the position space, rather than a position and coin register.  Thus the Hilbert space is spanned by $\ket {n,m},\ n,m=-\infty...\infty$.  We can describe the coin based walks using this Hilbert space by identifying $\ket{n,L}$ with $\ket{n,n+1}$ and $\ket{n,R}$ with $\ket{n,n-1}$.

In the case of motion on a line, a double-reflection walk is set up  as follows \cite{szegedy,magniez2007a}.  Let $d_n,e_n,f_n$ and $g_n$ be complex numbers satisfying
\begin{equation}
\begin{split}
|d_n|^2 + |e_n|^2 = 1, \forall n \in \mathbb{Z} \\
|f_n|^2 + |g_n|^2 = 1, \forall n \in \mathbb{Z}
\end{split}
\end{equation}

The two-step walk operator $W_{DR}$ is defined via the following equations:
\begin{equation}
\label{simple-pn-qm}
\begin{split}
\ket{p_n} & = d_n\ket{n+1} + e_n\ket{n-1} \\
\ket{q_m} & = f_m\ket{m+1} + g_m\ket{m-1}
\end{split}
\end{equation}
\begin{equation}
\begin{split}
\Pi_A & = \sum_n \ket{n}\ket{p_n} \bra{n}\bra{p_n} \\
\Pi_B & = \sum_m \ket{q_m}\ket{m} \bra{q_m}\bra{m}
\end{split}
\end{equation}

\begin{equation}
\begin{split}
\label{W_DR}
W_{DR} = (2\Pi_B-I)(2\Pi_A-I)
\end{split}
\end{equation}

The Hadamard walk in the double-reflection framework can be realized by setting:
\begin{equation}
\begin{split}
d_n & = f_n = \frac{1}{2}\sqrt{2+\sqrt{2}} \\
e_n & = g_n = \frac{1}{2}\sqrt{2-\sqrt{2}}
\end{split}
\end{equation}
The inhomogeneous quantum walk with coin in equation (\ref{eqn-cos_sin_matrix}) can be reproduced by taking:
\begin{equation}
\begin{split}
d_n & = f_n = \frac{1}{\sqrt{2}} \sqrt{1+\sin\left( \frac{n\pi}{k} \right)} \\
e_n & = g_n = \frac{1}{\sqrt{2}} \sqrt{1-\sin\left( \frac{n\pi}{k} \right)}
\end{split}
\end{equation}

Although the original walks in \cite{szegedy} have real amplitudes, it is natural to allow the constants $d_n,e_n,f_n$ and $g_n$ be complex numbers as in \cite{magniez2007a}.  We now show that even if we allow this,
not all walks produced by the position-dependant coin construction can be realized by walks of the form (\ref{W_DR}).

Using a generalized double-reflection form of the quantum walk to $\ket{\psi_0} = a_0 \ket{n,n+1} + b_0 \ket{n,n-1}$ yields:
\begin{equation}
\begin{split}
\label{W-S-psi-0}
& W_{DR}\ket{\psi_0} = \\ &   2\bar{f}_{n-1}g_{n-1} \left( 2a_0\bar{d}_n e_n + b_0\left(2|e_n|^2-1\right) \right) \ket{n-2,n-1} \\ + & \left( 2|f_{n-1}|^2-1 \right) \left( 2a_0\bar{d}_n e_n + b_0\left(2|e_n|^2-1\right) \right) \ket{n,n-1} \\ + & \left(2|g_{n+1}|^2-1 \right) \left( a_0\left(2|d_n|^2-1\right) + 2b_0d_n\bar{e}_n \right) \ket{n,n+1} \\ + & 2f_{n+1}\bar{g}_{n+1} \left( a_0\left(2|d_n|^2-1\right) + 2b_0 d_n\bar{e}_n \right) \ket{n+2,n+1}
\end{split}
\end{equation}

We now wish to compare this to a walk using a controlled unitary coin.  Since we have allowed the  \lq\lq forward\rq\rq\ and \lq\lq backward\rq\rq\ steps to be different
in (\ref{W-S-psi-0}), we need to take this into account.  So for two steps of the position-dependent-coin walk we use different coin operators in the
first and second step.

We consider a two general controlled unitary coins, described in the Hilbert space of two copies of the position register:
\begin{equation}
\begin{split}
U_n \ket{n,n+1} & = \alpha_n \ket{n,n+1} + \beta_n \ket{n,n-1} \\
U_n \ket{n,n-1} & = -e^{i\theta_n}\bar{\beta_n} \ket{n,n+1} + e^{i\theta_n}\bar{\alpha_n} \ket{n,n-1}
\end{split}
\end{equation}
and
\begin{equation}
\begin{split}
\tilde U_n \ket{n,n+1} & = \tilde \alpha_n \ket{n,n+1} + \tilde \beta_n \ket{n,n-1} \\
\tilde U_n \ket{n,n-1} & = -e^{i\tilde \theta_n}\overline{\tilde \beta}_n \ket{n,n+1} + e^{i\tilde \theta_n}\overline{\tilde \alpha}_n \ket{n,n-1}
\end{split}
\end{equation}

The general two-step  walk operator associated with these coins is given by:
\begin{equation}
W_{PDC} = S \tilde C S C
\end{equation}
where
\begin{equation}
C= \displaystyle \sum_{m=-\infty}^\infty \ket{m}\bra{m} \otimes U_m,
\end{equation}
\begin{equation}
\tilde C= \displaystyle \sum_{m=-\infty}^\infty \ket{m}\bra{m} \otimes \tilde U_m,
\end{equation}
and
\begin{equation}
\begin{split}
&S =  \sum_{m=-\infty}^\infty \ket{m-1,m}\bra{m,m+1} + \ket{m+1,m}\bra{m,m-1}\\
&\quad + \sum_{m\neq n+1\ {\rm and}\ m\neq n-1} \ket{mn}\bra{mn}
\end{split}
\end{equation}

Let $\ket{\psi_0} = a_0\ket{n,n+1} + b_0\ket{n,n-1}$ with $|a_0|^2 + |b_0|^2 = 1$. Applying  walk operator yields:
\begin{equation}
\label{W-PD-psi-0}
\begin{split}
& W_{PDC}\ket{\psi_0}=\\
&  \tilde\alpha_{n-1} \left( a_0\alpha_n - b_0e^{i\theta_n}\bar{\beta}_n \right) \ket{n-2,n-1} \\
& + \tilde\beta_{n-1} \left( a_0\alpha_n - b_0 e^{i\theta_n} \bar{\beta}_n \right) \ket{n,n-1} \\
& - e^{i\tilde\theta_{n+1}}\bar{\tilde\beta}_{n+1} \left( a_0 \beta_n + b_0 e^{i\theta_n} \bar{\alpha}_n \right) \ket{n,n+1} \\
& + e^{i\tilde\theta_{n+1}}\bar{\tilde\alpha}_{n+1} \left( a_0 \beta_n + b_0 e^{i\theta_n} \bar{\alpha}_n \right) \ket{n+2,n+1}
\end{split}
\end{equation}

We can now see that the effect of $W_{DR}$ can be
reproduced by a walk using a position-dependent coins by choosing
$\alpha_n = 2\bar{d}_n e_n$, $\beta_n = -\left(2|e_n|^2-1\right)$, $\theta_n=0$  and
$\tilde\alpha_n = 2\bar{f}_n g_n$, $\tilde\beta_n = -\left(2|g_n|^2-1\right)$, $\tilde\theta_n=0$.

However the most general walk of the form (\ref{W-S-psi-0}) cannot reproduce all walks with position dependent coins.  This is because from (\ref{W-S-psi-0}) one can see that
\begin{equation}
\bra{n,n+1}W_{DR} \ket{n,n+1}=\left(2|g_{n+1}|^2-1 \right) \left( 2|d_n|^2-1\right)
\end{equation}
is real for any choice of $d_n,e_n,f_n$ and $g_n$, however
\begin{equation}
\bra{n,n+1}W_{PDC} \ket{n,n+1} = -e^{i\tilde\theta_{n+1}}\bar{\tilde\beta}_{n+1}\beta_n,
\end{equation}
which need not be real.  Indeed even if we take both steps of the position-dependent-coin walk to be the same (as is usually done),
we get
\begin{equation}
\bra{n,n+1}W_{PDC} \ket{n,n+1} = -e^{i\theta_{n+1}}\bar{\beta}_{n+1}\beta_n,
\end{equation}
and still the most general walk of the form (\ref{W-S-psi-0}) cannot reproduce all walks with position dependent coins.

Indeed it is not too difficult to check that even if one allows transitions to arbitrary positions in the double-reflection form of the walk,
it cannot reproduce all walks defined by position-dependent coins.  For let us take  general states of the form
\begin{equation}
\label{general-pn-qm}
\begin{split}
\ket{P_n} & = \sum_{j=-\infty}^{\infty} D_{nj}\ket{n+j} \\
\ket{Q_m} & = \sum_{j=-\infty}^{\infty} E_{mj}\ket{m+j},
\end{split}
\end{equation}
rather than (\ref{simple-pn-qm}).  The matrix element
\begin{equation}
\begin{split}
& \bra{n,n+1}W_{DR} \ket{n,n+1} \\
& \quad=\big(2|\langle n \ket{Q_{n+1}}|^2-1\big) \big(2|\langle n+1 \ket{P_{n}}|^2-1\big)
\end{split}
\end{equation}
is still real.

We also note that for some purposes it is natural to use the shift operator 
\begin{equation}
S = \displaystyle \sum_n \ket{n-1,R}\bra{n,L} + \ket{n+1,L}\bra{n,R}
\end{equation}
rather than (\ref{standard-shift}).  This is a closer analogue of the shift embodied in the double-reflection framework.  It is straightforward
to check, however, that this does not alter our conclusions: this modified version of the position-dependent-coin walk can also reproduce the double-reflection
walk; and not all position-dependent-coin walks with this modified shift can be realized by double-reflection walks.

Nonetheless, there may be a generalization of the double-reflection form of the quantum walk which allows for these two constructions to be equivalent; this question is left open for future work.

\section{\label{sec:section4}The P\'{o}lya Urn}

The classical P\'{o}lya Urn is a model in which there is an urn containing a number of red and black balls. At every time step one is drawn at random, then replaced with a copy of the drawn ball. Let $R_n$ be a random variable representing the number of red balls in an urn containing $n$ balls in total. Let the random variable $X_n$ be given by $X_n = \frac{R_n}{n}$, then it is possible to show that $X_n \rightarrow \beta(r_0, b_0)$ almost surely as $n\rightarrow \infty$, where $r_0$ and $b_0$ are the initial number of red and black balls in the urn respectively, and $\beta(r_0, b_0)$ is a beta distribution with parameters $r_0$ and $b_0$. \cite{freedman1965}.


The classical system is not irreducible so it does not fit naturally into the double-reflection framework.  It is nonetheless instructive to
see what happens if we try to use that framework to quantize this system.

A natural position space is spanned by $\ket {r,b}$ where $r$ and $b$ are the numbers of red and black balls respectively.  In the classical model
a time step increases the number of red or black balls by one so it is natural to try to set up a quantum model by taking something like the following:
\begin{equation}
\label{dr-PolyaUrn}
\begin{split}
\ket{p_{r,b}} & = \alpha_{r,b}\ket{r+1,b} + \beta_{r,b}\ket{r,b+1}, \\
\ket{q_{r,b}} & = \gamma_{r,b}\ket{r+1,b} + \delta_{r,b}\ket{r,b+1},
\end{split}
\end{equation}
\begin{equation}
\begin{split}
\Pi_A & = \sum_{r,b} \ket{r,b}\bra{r,b}\otimes\ket{p_{r,b}}\bra{p_{r,b}}, \\
\Pi_B & = \sum_{r,b} \ket{q_{r,b}}\bra{q_{r,b}}\otimes\ket{r,b}\bra{r,b},
\end{split}
\end{equation}
\begin{equation}
\label{W_S}
W_{Polya-DR} = (2\Pi_B-I)(2\Pi_A-I),
\end{equation}
where
$\alpha_{r,b},\beta_{r,b},\gamma_{r,b},\delta_{r,b}$ are  constants.

However it may easily be checked that this system has the following behaviour
\begin{itemize}
\item for each pair $(r,b)$ the two-dimensional subspace spanned by $\ket {r,b}\ket {r+1,b}$ and $\ket {r,b}\ket {r,b+1}$
is invariant under the time evolution (so if the starting state of our walk is in this subspace, it just moves
around in the subspace)
\item similarly for each pair $(r,b)$ the two-dimensional subspace spanned by $\ket {r+1,b}\ket {r,b}$ and $\ket {r,b+1}\ket {r,b}$
is invariant under the time evolution
\item all other states are unchanged by the time evolution
\end{itemize}

In particular the number of red or black balls does not increase with time by more than one from its initial value.  Thus this quantum evolution does not seem to be the natural quantum version of the classical model.  It would be interesting to know whether it is possible to use  different states than (\ref{dr-PolyaUrn}) to produce a double-reflection quantization of the P\'{o}lya Urn that is more satisfactory (i.e. one that only increases the number of red and black balls as time evolves), but we have not yet been able to do so.

It turns out to be relatively straightforward to set up  a quantum walk with a position-dependent coin that seems to capture how a quantum version of the P\'{o}lya Urn should behave, as we now describe.

The state space is spanned by $\ket{r,b,R}$ and $\ket{r,b,B}$.  The first two registers are the number of red and black balls and the third
register can be in one of two states $R$ or $B$ corresponding to whether the number of red or black balls is to increase in the \lq\lq shift\rq\rq\ step of the walk.
We are really only interested in the number of red and black balls $r$ and $b$ being positive, but in order for the walk to be
unitary it will be convenient to allow $r$ and $b$ to take any integer values (although we will typically only be interested
in initial states with $r,b\ge 0$ and $r+b\ge 1$).
So we set up our walk as follows.  The coin operator is defined by:
\begin{equation}
C_{r,b}\ket{R}=
\begin{cases} \sqrt{\frac{r}{r+b}}\ket{R} + \sqrt{\frac{b}{r+b}}\ket{B} & \text{if $r,b,r+b-1\ge 0$,}
\\
\ket{R} &\text{otherwise.}
\end{cases}
\end{equation}
\begin{equation}
C_{r,b}\ket{B}=
\begin{cases} \sqrt{\frac{b}{r+b}}\ket{R} - \sqrt{\frac{r}{r+b}}\ket{B} & \text{if $r,b,r+b-1\ge 0$,}
\\
\ket{B} &\text{otherwise.}
\end{cases}
\end{equation}
\begin{equation}
\begin{split}
S & = \displaystyle \sum_{r,b=-\infty}^\infty \ket{r+1,b,R}\bra{r,b,R} + \ket{r,b+1,B}\bra{r,b,B}
\end{split}
\end{equation}

\begin{equation}
W_{Polya-PDC} = S \left( \displaystyle \sum_{r,b=-\infty}^\infty \ket{r,b}\bra{r,b} \otimes C_{r,b} \right)
\end{equation}


An attractive feature of this walk is that if you run one step of the walk, measure the system and reset the chirality state to $\ket{R}$, then the classical walk is recovered.

\begin{figure}[htb]
\centering
\includegraphics[keepaspectratio=false, width=3in, height=2in]{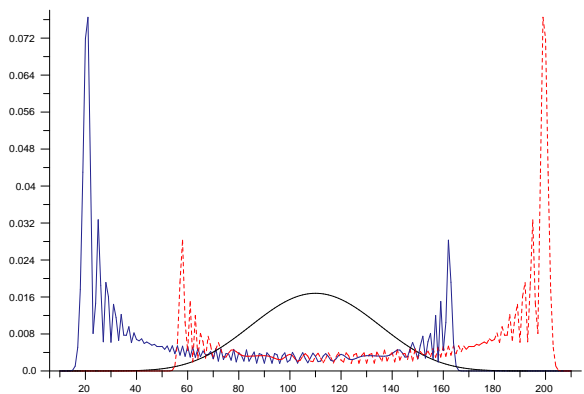}
\caption{Probability distributions for number of red balls of the quantum P\'{o}lya Urn (dashed red line), number of black balls of the quantum P\'{o}lya urn (solid black line) and the number of red balls of the classical P\'{o}lya Urn (dot-dashed blue line) with $r_0=b_0=10$ for 200 time steps.}\label{fig:polya_r10_b10_200t_biased}
\end{figure}

The resulting probability distribution of running this walk, starting with 10 red balls and 10 black balls for 200 time steps, can be seen in Figure \ref{fig:polya_r10_b10_200t_biased}.

There is a heavy bias to the right, similar to the Hadamard random walk. It is possible to reduce the bias using a more symmetric coin and initial condition.

The quantum P\'{o}lya urn plot shows several features that one might expect, such as most of the probability being concentrated to the far right or left. This is to be expected, as the P\'{o}lya urn is an example of a reinforced process; if an event occurs then it becomes more likely to occur in the future.

\begin{figure}[htb]
\centering
\includegraphics[keepaspectratio=false, width=3in, height=2in]{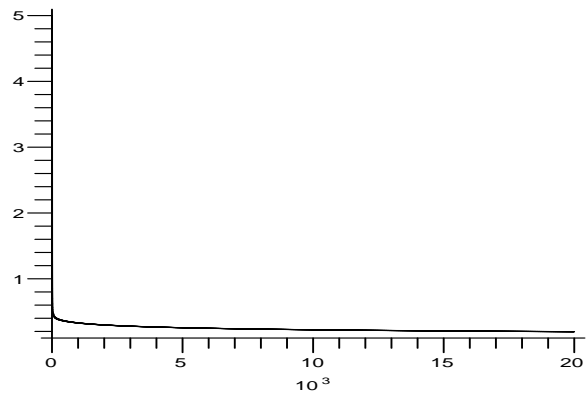}
\caption{Standard deviation divided by number of time steps for the quantum P\'{o}lya Urn with $r_0=10$ and $b_0=10$ run for 20000 time steps}\label{fig:polya_sd_r10_b10_20000t_n}
\end{figure}

We also calculated the standard deviation of the P\'{o}lya Urn quantum walk numerically, and a plot of this divided by the number of timesteps can be seen in Figure \ref{fig:polya_sd_r10_b10_20000t_n}. For the walk to evolve linearly in time asymptotically as $T\rightarrow \infty$, one would expect this graph to tend towards a horizontal line. This appears to be the case, but a proof of asymptotic linear evolution is an open question for future work as the quantum P\'{o}lya Urn has a non-periodic coin and thus our original method for periodic coins can not be applied.

\section{\label{sec:section5}Conclusions}

In this paper we have analyzed some particular models of inhomogeneous walks.  It would be interesting to understand whether one can make general
characterizations of the long-time behaviour of quantum walks from knowledge of their (position-dependent) coins. For example we have described walks
that are bounded and others whose spread is linear in time. We do not know what intermediate types of behaviour are possible.  It would also
be interesting to know whether quantum walks (including non-periodic ones) \lq\lq typically\rq\rq\ spread linearly in time.  It would
also be attractive to understand the relationship between the double-reflection and position-dependent-coin walks more fully: is there a
suitable generalization that encompasses both?

\begin{acknowledgments}
We are very grateful to Aram Harrow, Miklos Santha and Andreas Winter for many insightful observations.
We also gratefully acknowledge support for this work from:
the UK EPSRC through the
QIP-IRC,  the University of Bristol for a Research
Fellowship and  the EU through the
project QAP.
\end{acknowledgments}

\newpage
\bibliography{IQW-19jun09}

\begin{thebibliography}{17}
\expandafter\ifx\csname natexlab\endcsname\relax\def\natexlab#1{#1}\fi
\expandafter\ifx\csname bibnamefont\endcsname\relax
  \def\bibnamefont#1{#1}\fi
\expandafter\ifx\csname bibfnamefont\endcsname\relax
  \def\bibfnamefont#1{#1}\fi
\expandafter\ifx\csname citenamefont\endcsname\relax
  \def\citenamefont#1{#1}\fi
\expandafter\ifx\csname url\endcsname\relax
  \def\url#1{\texttt{#1}}\fi
\expandafter\ifx\csname urlprefix\endcsname\relax\def\urlprefix{URL }\fi
\providecommand{\bibinfo}[2]{#2}
\providecommand{\eprint}[2][]{\url{#2}}

\bibitem[{\citenamefont{Aharonov et~al.}(1993)\citenamefont{Aharonov,
  Davidovich, and Zagury}}]{aharonov1993}
\bibinfo{author}{\bibfnamefont{Y.}~\bibnamefont{Aharonov}},
  \bibinfo{author}{\bibfnamefont{L.}~\bibnamefont{Davidovich}},
  \bibnamefont{and} \bibinfo{author}{\bibfnamefont{N.}~\bibnamefont{Zagury}},
  \bibinfo{journal}{Phys. Rev. A} \textbf{\bibinfo{volume}{48}},
  \bibinfo{pages}{1687} (\bibinfo{year}{1993}).

\bibitem[{\citenamefont{Meyer}(1996{\natexlab{a}})}]{meyer1996a}
\bibinfo{author}{\bibfnamefont{D.}~\bibnamefont{Meyer}}, \bibinfo{journal}{J.
  Stat. Phys.} \textbf{\bibinfo{volume}{85}}, \bibinfo{pages}{551}
  (\bibinfo{year}{1996}{\natexlab{a}}).

\bibitem[{\citenamefont{Meyer}(1996{\natexlab{b}})}]{meyer1996b}
\bibinfo{author}{\bibfnamefont{D.}~\bibnamefont{Meyer}},
  \bibinfo{journal}{Phys. Lett. A} \textbf{\bibinfo{volume}{223}},
  \bibinfo{pages}{337} (\bibinfo{year}{1996}{\natexlab{b}}).

\bibitem[{\citenamefont{Watrous}(2001)}]{watrous2001}
\bibinfo{author}{\bibfnamefont{J.}~\bibnamefont{Watrous}},
  \bibinfo{journal}{Journal of Computer and System Sciences}
  \textbf{\bibinfo{volume}{62}}, \bibinfo{pages}{376} (\bibinfo{year}{2001}).

\bibitem[{\citenamefont{Shenvi et~al.}(2003)\citenamefont{Shenvi, J.Kempe, and
  Whaley}}]{shenvi2003}
\bibinfo{author}{\bibfnamefont{N.}~\bibnamefont{Shenvi}},
  \bibinfo{author}{\bibnamefont{J.Kempe}}, \bibnamefont{and}
  \bibinfo{author}{\bibfnamefont{K.}~\bibnamefont{Whaley}},
  \bibinfo{journal}{Physical Review A} \textbf{\bibinfo{volume}{67}}
  (\bibinfo{year}{2003}).

\bibitem[{\citenamefont{Ambainis}(2007)}]{ambainis-2007-37}
\bibinfo{author}{\bibfnamefont{A.}~\bibnamefont{Ambainis}},
  \bibinfo{journal}{SIAM Journal on Computing} \textbf{\bibinfo{volume}{37}},
  \bibinfo{pages}{210} (\bibinfo{year}{2007}).

\bibitem[{\citenamefont{Buhrman and Spalek}(2006)}]{buhrman2004}
\bibinfo{author}{\bibfnamefont{H.}~\bibnamefont{Buhrman}} \bibnamefont{and}
  \bibinfo{author}{\bibfnamefont{R.}~\bibnamefont{Spalek}},
  \bibinfo{journal}{Proc. 17th ACM-SIAM Symposium on Discrete Algorithms} p.
  \bibinfo{pages}{880} (\bibinfo{year}{2006}).

\bibitem[{\citenamefont{Magniez and Nayak}(2007)}]{magniez-nayak2007}
\bibinfo{author}{\bibfnamefont{F.}~\bibnamefont{Magniez}} \bibnamefont{and}
  \bibinfo{author}{\bibfnamefont{A.}~\bibnamefont{Nayak}},
  \bibinfo{journal}{Algorithmica} \textbf{\bibinfo{volume}{48}},
  \bibinfo{pages}{221} (\bibinfo{year}{2007}).

\bibitem[{\citenamefont{Magniez
  et~al.}(2007{\natexlab{a}})\citenamefont{Magniez, Santha, and
  Szegedy}}]{magniez2007b}
\bibinfo{author}{\bibfnamefont{F.}~\bibnamefont{Magniez}},
  \bibinfo{author}{\bibfnamefont{M.}~\bibnamefont{Santha}}, \bibnamefont{and}
  \bibinfo{author}{\bibfnamefont{M.}~\bibnamefont{Szegedy}},
  \bibinfo{journal}{SIAM Journal of Computing} \textbf{\bibinfo{volume}{37}},
  \bibinfo{pages}{413} (\bibinfo{year}{2007}{\natexlab{a}}).

\bibitem[{\citenamefont{Ambainis}(2003)}]{ambainis2003}
\bibinfo{author}{\bibfnamefont{A.}~\bibnamefont{Ambainis}},
  \bibinfo{journal}{International Journal of Quantum Information}
  \textbf{\bibinfo{volume}{1}}, \bibinfo{pages}{507} (\bibinfo{year}{2003}).

\bibitem[{\citenamefont{Ambainis et~al.}(2005)\citenamefont{Ambainis, Kempe,
  and Rivosh}}]{ambainis2005}
\bibinfo{author}{\bibfnamefont{A.}~\bibnamefont{Ambainis}},
  \bibinfo{author}{\bibfnamefont{J.}~\bibnamefont{Kempe}}, \bibnamefont{and}
  \bibinfo{author}{\bibfnamefont{A.}~\bibnamefont{Rivosh}},
  \bibinfo{journal}{Proc. 16th ACM-SIAM SODA} p. \bibinfo{pages}{1099}
  (\bibinfo{year}{2005}).

\bibitem[{\citenamefont{Santha}(2008)}]{santha2008}
\bibinfo{author}{\bibfnamefont{M.}~\bibnamefont{Santha}}, \bibinfo{journal}{5th
  Theory and Applications of Models of Computation (TAMC08), Xian, April 2008,
  LNCS 4978} p.~\bibinfo{pages}{31} (\bibinfo{year}{2008}).

\bibitem[{\citenamefont{Szegedy}(2004)}]{szegedy}
\bibinfo{author}{\bibfnamefont{M.}~\bibnamefont{Szegedy}}, in
  \emph{\bibinfo{booktitle}{FOCS '04: Proceedings of the 45th Annual IEEE
  Symposium on Foundations of Computer Science}} (\bibinfo{publisher}{IEEE
  Computer Society}, \bibinfo{address}{Washington, DC, USA},
  \bibinfo{year}{2004}), pp. \bibinfo{pages}{32--41}, ISBN
  \bibinfo{isbn}{0-7695-2228-9}.

\bibitem[{\citenamefont{Magniez
  et~al.}(2007{\natexlab{b}})\citenamefont{Magniez, Nayak, Roland, and
  Santha}}]{magniez2007a}
\bibinfo{author}{\bibfnamefont{F.}~\bibnamefont{Magniez}},
  \bibinfo{author}{\bibfnamefont{A.}~\bibnamefont{Nayak}},
  \bibinfo{author}{\bibfnamefont{J.}~\bibnamefont{Roland}}, \bibnamefont{and}
  \bibinfo{author}{\bibfnamefont{M.}~\bibnamefont{Santha}},
  \bibinfo{journal}{Proc. 39th ACM Symposium on the Theory of Computing} p.
  \bibinfo{pages}{575} (\bibinfo{year}{2007}{\natexlab{b}}).

\bibitem[{\citenamefont{Nayak and Vishwanath}(2000)}]{nayak-2000}
\bibinfo{author}{\bibfnamefont{A.}~\bibnamefont{Nayak}} \bibnamefont{and}
  \bibinfo{author}{\bibfnamefont{A.}~\bibnamefont{Vishwanath}},
  \bibinfo{journal}{arXiv:quant-ph/0010117}  (\bibinfo{year}{2000}).

\bibitem[{\citenamefont{Ambainis et~al.}(2001)\citenamefont{Ambainis, Bach,
  Nayak, Vishwanath, and Watrous}}]{ambainis2001}
\bibinfo{author}{\bibfnamefont{A.}~\bibnamefont{Ambainis}},
  \bibinfo{author}{\bibfnamefont{E.}~\bibnamefont{Bach}},
  \bibinfo{author}{\bibfnamefont{A.}~\bibnamefont{Nayak}},
  \bibinfo{author}{\bibfnamefont{A.}~\bibnamefont{Vishwanath}},
  \bibnamefont{and} \bibinfo{author}{\bibfnamefont{J.}~\bibnamefont{Watrous}},
  \bibinfo{journal}{Proceedings of the 33rd Annual AMC Symposium on Theory of
  Computing} pp. \bibinfo{pages}{37--49} (\bibinfo{year}{2001}).

\bibitem[{\citenamefont{Freedman}(1965)}]{freedman1965}
\bibinfo{author}{\bibfnamefont{D.~A.} \bibnamefont{Freedman}},
  \bibinfo{journal}{Ann. Math. Statist.} \textbf{\bibinfo{volume}{36}},
  \bibinfo{pages}{956} (\bibinfo{year}{1965}).

\end{thebibliography}

\end{document}